\documentclass[aps,prl,twocolumn,showpacs,superscriptaddress,groupedaddress,nofootinbib]{revtex4}
\usepackage{amsmath,amsfonts,amssymb,array}
\usepackage{bm,bbm,bbm,bbold}
\usepackage{enumerate}
\usepackage{dsfont}
\usepackage{float}
\usepackage{graphicx}
\usepackage{longtable}
\usepackage{multirow}
\usepackage{slashed,subfigure}
\usepackage{xcolor}
\usepackage[colorlinks = true,linkcolor = blue]{hyperref} 
\providecommand{\tabularnewline}{\\}

\def\DYCC{{\rm CC~DY}}
\def\DYNC{{\rm NC~DY}}
\def\DYj{{\rm DY}+1j}

\def\VBF{{\rm VBF}}
\def\GF{{\rm GF}}

\def\fb{{\rm ~fb}}
\def\pb{{\rm ~pb}}

\def\GeV{{\rm ~GeV}}
\def\TeV{{\rm ~TeV}}

\newcommand{\mgamc}{MadGraph5\_aMC@NLO}
\newcommand{\nloct}{N{\small LO}CT}
\begin{document}
\leftline{}
\rightline{IPPP-16-13, MCnet-16-05}

\title{Fully Automated Precision Predictions for\\ Heavy Neutrino Production Mechanisms at Hadron Colliders}

\author{C\'eline Degrande}  
\email{celine.degrande@durham.ac.uk}
\affiliation{Institute for Particle Physics Phenomenology, Department of Physics, Durham University, Durham DH1 3LE, U.K.}

\author{Olivier Mattelaer}  
\email{o.p.c.mattelaer@durham.ac.uk}
\affiliation{Institute for Particle Physics Phenomenology, Department of Physics, Durham University, Durham DH1 3LE, U.K.}

\author{Richard Ruiz}  
\email{richard.ruiz@durham.ac.uk}
\affiliation{Institute for Particle Physics Phenomenology, Department of Physics, Durham University, Durham DH1 3LE, U.K.}

\author{Jessica Turner}  
\email{jessica.turner@durham.ac.uk}
\affiliation{Institute for Particle Physics Phenomenology, Department of Physics, Durham University, Durham DH1 3LE, U.K.}

\date{\today}

\begin{abstract}
Motivated by TeV-scale neutrino mass models, we propose a systematic treatment of heavy neutrino $(N)$ production at hadron colliders.
Our simple and efficient modeling of the vector boson fusion (VBF) $W^\pm\gamma\rightarrow N\ell^\pm$ and $N\ell^\pm+nj$ signal definitions
resolve collinear and soft divergences that have plagued past studies, and is applicable to other color-singlet processes, 
e.g., associated Higgs $(W^\pm h)$,  sparticle $(\tilde{\ell}^\pm\tilde{\nu_\ell})$, and charged Higgs $(h^{\pm\pm}h^{\mp})$ production.
We present, for the first time, a comparison of all leading $N$ production modes, 
including both gluon fusion (GF) $gg\rightarrow Z^*/h^*\rightarrow N\overset{(-)}{\nu_\ell}$ and VBF.
We obtain fully differential results up to next-to-leading order (NLO) in QCD accuracy 
using a Monte Carlo tool chain linking FeynRules, N{\small LO}CT, and MadGraph5\_aMC@NLO.
Associated model files are publicly available.
At the 14 TeV LHC, the leading order GF rate is small and comparable to the NLO $N\ell^\pm+1j$ rate;
at a future 100 TeV Very Large Hadron Collider, GF dominates for $m_N={300-1500}$ GeV, beyond which VBF takes lead.
\end{abstract}


\maketitle

\section{Introduction}\label{sec:Intro}
The origin of neutrino masses $m_\nu$ that are tiny compared to all other fermion masses is a broad issue
in particle physics, cosmology, and astrophysics.
Nonzero $m_\nu$ imply the existence of new particles~\cite{Ma:1998dn},
but more generally, physics beyond the Standard Model (BSM) that may be observable at current and future experiments.
Extended neutrino mass 
models~\cite{Pilaftsis:1991ug,Kersten:2007vk,Gu:2008yj,Dev:2009aw,Zhang:2009ac,Adhikari:2010yt,Chen:2013fna,Dev:2013oxa,Dev:2015vra} based on the
Type I~\cite{Minkowski:1977sc,Yanagida:1979as,VanNieuwenhuizen:1979hm,Ramond:1979py,Glashow:1979nm,Mohapatra:1979ia,
GellMann:1980vs,Schechter:1980gr,Shrock:1980ct,Schechter:1981cv}, 
Inverse~\cite{Mohapatra:1986aw,Mohapatra:1986bd,Bernabeu:1987gr}, and 
Linear Seesaw Mechanisms~\cite{Akhmedov:1995ip,Akhmedov:1995vm}, 
feature heavy mass eigenstates $N_{i}$ that couple to electroweak (EW) bosons via mixing with left-handed (LH) neutrinos $\nu_{L}$.
In these TeV-scale scenarios, active-sterile mixing can be as large as  $\vert V_{\ell N_i}\vert\sim10^{-3}-10^{-2}$, 
and consistent with oscillation and EW data~\cite{Atre:2009rg,Antusch:2014woa,deGouvea:2015euy},
as well as direct searches by the Large Hadron Collider (LHC) experiments~\cite{Aaij:2014aba,Aad:2015xaa,Khachatryan:2016olu}.
Thus, if kinematically accessible, 
hadron colliders can produce heavy neutrinos that
decay to lepton number- and/or flavor-violating final states with observable rates.

\begin{figure*}[!th]
\includegraphics[width=0.90\textwidth]{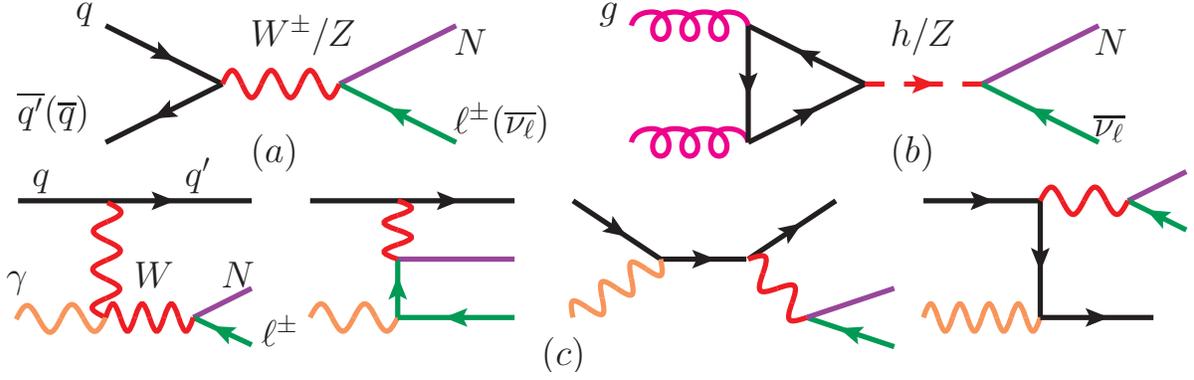}
\caption{Heavy neutrino production via (a) charge (neutral) current Drell-Yan, (b) gluon fusion, and (c) $W\gamma$ fusion.}
\label{fig:diagrams}
\end{figure*}

For heavy $N$ masses $m_N$ above the EW scale, a systematic comparison of all leading single $N$ production modes cataloged 
in~\cite{Keung:1983uu,Datta:1993nm} has never been performed.
Most investigations focus on the charge current (CC) Drell-Yan (DY) 
process~\cite{Keung:1983uu,Pilaftsis:1991ug,Han:2006ip,delAguila:2006bda,delAguila:2007qnc,Atre:2009rg},
as shown in Fig.~\ref{fig:diagrams}(a),
\begin{equation}
 q ~\overline{q'} \rightarrow W^* \rightarrow N ~\ell^\pm, ~\quad q \in \{u,c,d,s,b\},
 \label{eq:DYCC}
\end{equation}
and has recently been found to be subleading in parts of this mass regime~\cite{Hessler:2014ssa,Alva:2014gxa}.
Missing in most analyses is the gluon fusion (GF) channel~\cite{Hessler:2014ssa},
which proceeds at leading order (LO) through quark triangles in Fig.~\ref{fig:diagrams}(b),
\begin{equation}
 g ~g \rightarrow h^*/Z^* \rightarrow N ~ \overset{(-)}{\nu_\ell}.
 \label{eq:GF}
\end{equation}
Variants of Eq.~(\ref{eq:GF}) have been studied elsewhere~\cite{BhupalDev:2012zg,Batell:2015aha}.
Formally, GF is a finite next-to-next-to-leading order in QCD correction to the neutral current (NC) DY process
\begin{eqnarray}
 q ~\overline{q} \rightarrow Z^* \rightarrow N ~\overset{(-)}{\nu_\ell}.
 \label{eq:DYNC}
\end{eqnarray}

Recent analyses have investigated the sizable EW vector boson fusion (VBF) 
process~\cite{Dev:2013wba,Bambhaniya:2014hla,Alva:2014gxa,Deppisch:2015qwa,Ng:2015hba,Arganda:2015ija,Das:2015toa}:
\begin{equation}
 q_{1} ~q_{2} ~\overset{W\gamma+WZ~\text{Fusion}}{\longrightarrow} ~N ~\ell^\pm ~q_{1}' ~q_{2}',
 \label{eq:VBF}
\end{equation}
and subleading CC DY with $n\geq1$ QCD jets~\cite{Dev:2013wba,Das:2014jxa,Das:2015toa}:
\begin{equation}
 p ~p \rightarrow W^{*}+ nj \rightarrow N\ell^\pm+  nj, \quad p,j\in\{\overset{(-)}{q},g\},
 \label{eq:DYnj}
\end{equation}
but with conflicting results.
The last two processes are plagued by soft and collinear poles in $s$- and $t$-channel exchanges of massless gauge bosons, 
issues usually associated with perturbative QCD, and require care~\cite{Alva:2014gxa,Ruiz:2015zca}.
E.g., inadequately regulated diverges are responsible for the overestimated 
cross sections claimed in~\cite{Dev:2013wba,Das:2014jxa,Deppisch:2015qwa,Das:2015toa}

We introduce a treatment that resolves all these issues.
Our results have widespread implications for SM and BSM physics:
the prescriptions for Eqs.~(\ref{eq:VBF}) and (\ref{eq:DYnj}) are applicable to, among other processes,
associated Higgs $(W^\pm h)$,  sparticle $(\tilde{\ell}^\pm\tilde{\nu_\ell})$, and charged Higgs $(h^{\pm\pm}h^{\mp})$ production.
To date, our study is the most accurate and comprehensive presentation of heavy $N$ production mechanisms at colliders.
It represents the first time properties of infrared and collinear (IRC) safety have been so rigorously imposed in this context, particularly to VBF.
Furthermore, we obtain modest next-to-leading order (NLO) in QCD corrections, demonstrating the stability of our approach.

We guarantee the perturbativity of the VBF process by factorizing and resumming the $t$-channel $\gamma$
into a DGLAP-evolved parton distribution function (PDF).
Using a $\gamma$-PDF, one considers instead, as shown in Fig.~\ref{fig:diagrams}(c),
\begin{equation}
 q ~\gamma \rightarrow N ~\ell^\pm ~q'.
\end{equation}
$WZ$ fusion is subleading and can be neglected~\cite{Alva:2014gxa}.
We regularize Eq.~(\ref{eq:DYnj}) by imposing transverse momentum $(p_T)$ cuts consistent with Collins-Soper-Sterman (CSS) 
$p_T$-resummation~\cite{Collins:1984kg}.
Our Monte Carlo (MC) framework allows us to compute fully differential Feynman diagrams up to one loop, 
and therefore GF at LO and the remaining processes at NLO;
only Eq.~(\ref{eq:DYCC}) has been evaluated before at NLO~\cite{Ruiz:2015gsa,Das:2016hof}.

At the 14 TeV LHC, the CC DY channel prevails for $N$ masses $m_N=150-850\GeV$; above this, the VBF cross section is larger.
However, due to the $gg$ luminosity increase, GF is the leading mechanism at a hypothetical future 100 TeV Very Large Hadron Collider (VLHC)
for $m_N = {300-1500}\GeV$; at higher $m_N$, VBF dominates.

We now introduce our theoretical model, computation procedure,
and signal definition prescription.
After presenting and discussing results, we conclude.

\section{Heavy Neutrino Model}\label{sec:Theory}
For  
$i~(m) =1,\dots,3$, LH (light) states and 
$j~(m')=1,\dots,n$, right-handed (heavy) states,
chiral neutrinos can be expanded into mass eigenstates by the rotation
\begin{eqnarray}
\begin{pmatrix} \nu_{Li} \\  N_{Rj}^c \end{pmatrix}
=
\begin{pmatrix} 
U_{3\times3} && V_{3\times n} \\
X_{n\times3} && Y_{n\times n}
\end{pmatrix}
\begin{pmatrix} \nu_{m} \\ N_{m'}^c \end{pmatrix}.
\end{eqnarray}
After rotating the charged leptons into the mass basis, 
which we take to be the identity matrix for simplicity,
$U_{3 \times3}$ is the observed light neutrino mixing matrix and
$V_{3 \times n}$ parameterizes active-heavy mixing. 
In the notation of~\cite{Atre:2009rg}, the flavor state $\nu_{\ell}$ in the mass basis is
\begin{equation}
 \nu_{\ell} = \sum_{m=1}^{3} U_{\ell m}\nu_{m} + \sum_{m'=1}^{n}V_{\ell m'} N^{c}_{m'}.
\end{equation}
For simplicity, we consider only one heavy mass eigenstate, labeled by $N$. This does not affect our conclusions.
The interaction Lagrangian with EW bosons is then
\begin{eqnarray}
  \mathcal{L}_{\rm Int.} = 
  &-& \frac{g}{\sqrt{2}}W^+_\mu		\sum_{\ell=e}^\tau \sum_{m=1}^3 ~\overline{\nu_m} ~U_{\ell m}^*	~\gamma^\mu P_L\ell^-\nonumber\\
  &-& \frac{g}{\sqrt{2}}W^+_\mu		\sum_{\ell=e}^\tau 		~\overline{N^c} ~V_{\ell N}^* 	~\gamma^\mu P_L\ell^-\nonumber\\
  &-& \frac{g}{2\cos\theta_W}Z_\mu	\sum_{\ell=e}^\tau \sum_{m=1}^3	~\overline{\nu_m} ~U_{\ell m}^*	~\gamma^\mu P_L\nu_\ell\nonumber\\
  &-& \frac{g}{2\cos\theta_W}Z_\mu	\sum_{\ell=e}^\tau 		~\overline{N^c} ~V_{\ell N}^*	~\gamma^\mu P_L\nu_\ell\nonumber\\
  &-& \frac{g m_N}{2M_W} h	\sum_{\ell=e}^\tau 		~\overline{N^c} ~V_{\ell N}^*	 P_L\nu_\ell + \text{H.c.}
  \label{eq:Lagrangian}
\end{eqnarray}

Precise values of $V_{\ell N}$ are model-dependent and are constrained by oscillation and collider experiments, 
tests of lepton universality, and $0\nu\beta\beta$-decay~\cite{Atre:2009rg,Antusch:2014woa,deGouvea:2015euy}.
However, $V_{\ell N}$ factorize in $N$ production cross sections such that
\begin{equation}
 \sigma(pp\rightarrow N~X) = \vert V_{N\ell}\vert^2 ~\times~ \sigma_{0}(pp\rightarrow N~X),
\end{equation}
where $\sigma_{0}$ is a model-independent ``bare'' cross section in which one sets $\vert V_{N\ell}\vert=1$.
Hence, our results are applicable to various heavy neutrino models.

\section{Computational Setup}\label{sec:MC}
We implement the above Lagrangian with Goldstone boson couplings in the Feynman gauge into FeynRules (FR) 2.3.10~\cite{Alloul:2013bka,Christensen:2008py}.
QCD renormalization and $R_2$ rational counterterms are calculated with \nloct~1.02 (prepackaged in FR)~\cite{Degrande:2014vpa} 
and FeynArts 3.8~\cite{Hahn:2000kx}.
Feynman rules are collected into a universal output file (UFO)~\cite{Degrande:2011ua} and is available publicly~\cite{nloFRModel}.
We obtain fully differential results using MadGraph5$\_$aMC$@$NLO  2.3.3~\cite{Alwall:2014hca}.
SM inputs are taken from the 2014 Particle Data Group~\cite{Agashe:2014kda}:
\begin{eqnarray}
 \alpha^{\rm \overline{MS}}(M_{Z})		&=& 1/127.940, \quad M_{Z}=91.1876\GeV, \nonumber\\ 
 \sin^{2}_{\rm \overline{MS}}(\theta_{W}) 	&=& 0.23126.
 \label{eq:smInputs}
\end{eqnarray}
We assume five massless quarks, take the Cabbibo-Kobayashi-Masakawa (CKM) matrix to be diagonal with unit entries,
and use the NLO NNPDF2.3 QED PDF (\texttt{lhaid:244600}) ~\cite{Ball:2013hta},
which features a $\gamma$-PDF with both elastic and inelastic components,
at collider energies of $\sqrt{s}=14$ and $100$ TeV. 
We extract $\alpha_s(\mu_r^2)$ from the PDFs.

\section{Infrared- and Collinear-Safe Hadron Collider Signal Definitions}\label{sec:SigDef}
To consistently compare channels and colliders, 
we follow the 2013 Snowmass recommendations~\cite{Avetisyan:2013onh}
and evaluate cross sections assuming the same fiducial acceptance.
In practice, however, one tunes cuts to specific colliders and final states.
Jet and charged lepton pseudorapidities $(\eta^{j,\ell})$ and charged lepton $p_T$ are required to satisfy~\cite{Avetisyan:2013onh}:
\begin{equation}
 \vert\eta^{j,\ell}\vert < 2.5, \quad p_T^\ell > 20\GeV.
 \label{eq:LepCuts}
\end{equation}
QCD radiation in Eq.~(\ref{eq:DYnj}) gives rise to fixed order (FO) cross sections that scale as powers of $\log(Q^2/q_T^2)$:
\begin{equation}
 \sigma(pp\rightarrow N\ell^\pm+nj) \sim \sum^n_k \alpha_s^k(Q^2)\log^{(2k-1)}\left(\frac{Q^2}{q_T^2}\right).
 \label{eq:qcdPowerSeries}
\end{equation}
$Q\sim m_N$ is the scale of the hard scattering process
and $q_T \equiv \sum_k^n p_{T,k}^{j}$ is the $(N\ell)$-system's transverse momentum,
which equals the sum of all jet $p_T$.
The perturbativity of these logarithms for TeV-scale leptons was studied in~\cite{Ruiz:2015zca}.
In the CSS $p_T$-resummation formalism~\cite{Collins:1984kg}, FO results are trustworthy when
$\alpha_s(Q^2)$ is perturbative, with $\Lambda_{\rm QCD} = 0.2\GeV$, and $q_T$ is comparable to $Q$:
\begin{equation}
 \log\frac{Q}{\Lambda_{\rm QCD}}\gg1 \quad\text{and}\quad \log^2\frac{Q}{q_T}\lesssim\log\frac{Q}{\Lambda_{\rm QCD}}.
\end{equation}
Imposing $Q = m_N$, jets in Eq.~(\ref{eq:qcdPowerSeries}) must satisfy
\begin{equation}
 \sum_k^n p_{T,k}^{j} \gtrsim m_N \times e^{-\sqrt{\log(m_N/\Lambda_{\rm QCD})}}.
 \label{eq:cssPTBound}
\end{equation}
Taking for example $n=1$ and $m_N$ up to $1~(1.5)\TeV$, the mass range of interest at 14 (100) TeV, this translates to
\begin{equation}
 p_T^j \gtrsim 55~(80)\GeV.
\end{equation}
Weaker $p_T^j$ cuts lead to artificially large logarithms and overestimated cross sections in Eq.~(\ref{eq:qcdPowerSeries}).
We cluster jets with FastJet~\cite{Cacciari:2005hq,Cacciari:2011ma} 
using the anti-$k_T$ algorithm~\cite{Cacciari:2008gp} with a separation parameter of $ \Delta R = 0.4$.
For differential events, we parton shower (PS) with Pythia 8.212~\cite{Sjostrand:2014zea}.

We equate the factorization and renormalization scales
to half the sum over final-state transverse masses (\texttt{dynamical\_scale\_choice=3} in \mgamc):
\begin{eqnarray}
 \mu_f, \mu_r, \mu_0 = \sum_{k=N,\ell,\text{jets}}\cfrac{m_{T,k}}{2} = \frac12\sum_{k} \sqrt{m_k^2 + p_{T,k}^2}.
\end{eqnarray}
We quantify the scale dependence by varying it over
\begin{equation}
 0.5  \leq \mu/ \mu_0 \leq2.
\end{equation}

In our framework, the CC DY rate at NLO can be calculated via the \mgamc~commands:
\begin{verbatim}
> import model HeavyN_NLO
> define p  = u c d s b u~ c~ d~ s~ b~ g
> define j  = p
> define mu = mu+ mu-
> generate p p > n2 mu [QCD]
> output PP_Nl_NLO; launch;
\end{verbatim}
Similarly, the inclusive NC DY at NLO is calculated by
\begin{verbatim}
> define vv = vm vm~
> generate p p > n2 vv [QCD]
> output PP_Nv_NLO; launch;
\end{verbatim}
and the inclusive CC $\DYj$ NLO rate by
\begin{verbatim}
> generate p p > n2 mu j QED=2 QCD=1 [QCD]
> output PP_Nl1j_NLO; launch;
\end{verbatim}

GF is a loop-induced process, which are only recently~\cite{Hirschi:2015iia} supported by \mgamc.
Such fully automated computations at NLO are unavailable since the two-loop technology does not currently exist.
We therefore perform the LO calculation matched and merged with up to one additional jet via the MLM scheme \cite{Mangano:2006rw}.
We discard loops that are actually virtual corrections to the DY process
and keep only diagrams where gluons do not appears in the loop.
The inclusive, unmatched LO GF rate can be calculated with
 \begin{verbatim}
> generate g g > n2 vv [QCD]
> output GGF_Nv_LO; launch;
\end{verbatim}
{Note that the $h^*/Z^*$ interference vanishes due to $C$-invariance/Fury's theorem
and the (anti-) symmetric nature of the residual $h~(Z)$ coupling~\cite{Willenbrock:1985tj,Hessler:2014ssa}.}

\begin{figure*}[!th]
  \subfigure[]{	\includegraphics[width=0.47\textwidth]{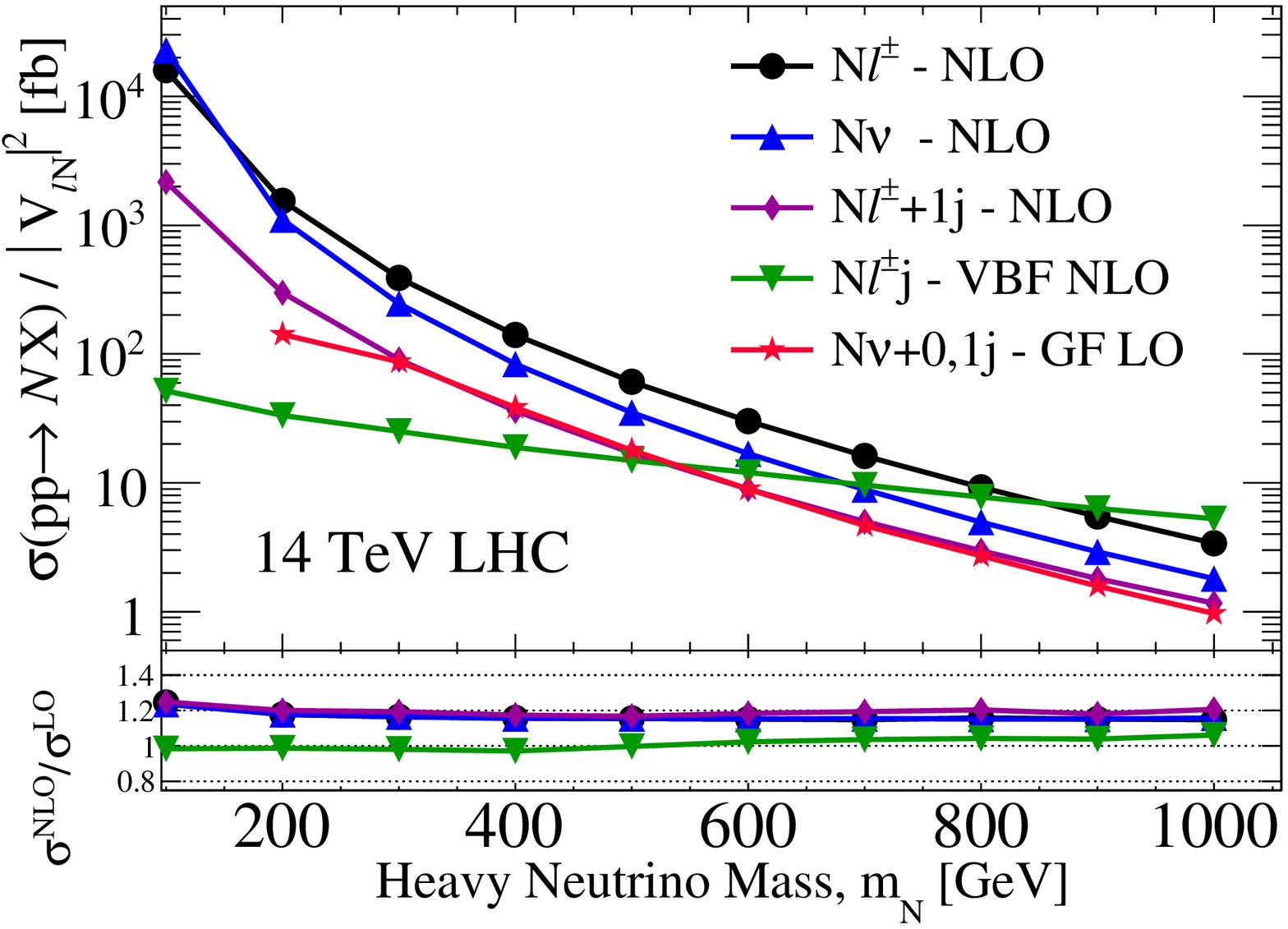}	\label{fig:xsec14TeV_mN}	}
  \subfigure[]{	\includegraphics[width=0.47\textwidth]{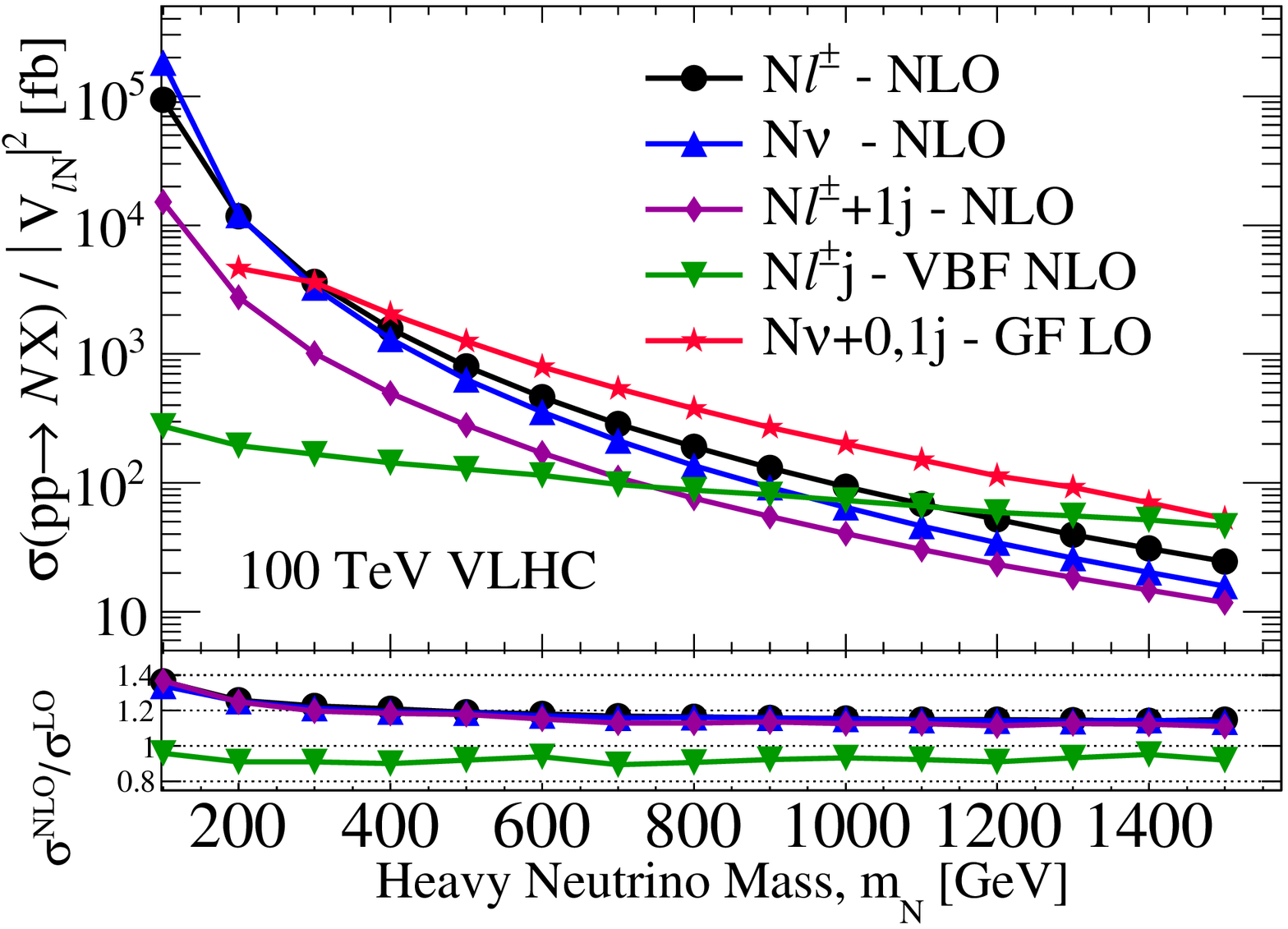}	\label{fig:xsec100TeV_mN}	}
\caption{
Heavy $N$ NLO production rate in (a) 14 and (b) 100 TeV $pp$ collisions as a function of $m_N$, 
divided by active-heavy mixing $\vert V_{\ell N}\vert^2$,
for the inclusive CC (circle) and NC (triangle) DY, $N\ell^\pm+1j$ (diamond), and VBF (upside-down triangle) processes, 
as well as the LO GF process matched up to $1j$ (star). Lower: Ratio of NLO and LO rates.}
\label{fig:xSec}
\end{figure*}

The difficulty in modeling $W\gamma$ fusion stems from the $t$-channel photon propagator, 
which, like Eq.~(\ref{eq:qcdPowerSeries}) for $N\ell^\pm+nj$, gives rise to logarithms of the form~\cite{Alva:2014gxa}
\begin{equation}
d\sigma(q_{1}q_{2}\rightarrow N \ell^\pm q_{1}' q_{2}') \sim 
\log\left(\cfrac{m_N^2}{M_W^2}\right)\log\left(\cfrac{m_N^2}{p_T^{j_\gamma 2}}\right).
\label{eq:VBFLogs}
\end{equation}
Here, $p_T^{j_\gamma}$ is the $p_T$ of the jet associated with the photon exchange.
However, consistent treatment of Eq.~(\ref{eq:VBFLogs}) dictates $p_T^j$ cuts excessive for $\gamma$-initiated processes. 
A resolution is to collinearly factorize and resum the photon piece into a DGLAP-evolved $\gamma$-PDF, consider instead
\begin{equation}
 q ~\gamma \rightarrow N ~\ell^\pm ~q',
 \label{eq:qaScattering}
\end{equation}
and evolve the PDF to the hard scattering scale.
One loses the ability to efficiently tag a second forward/backward jet but gains a large (logarithmic) total rate enhancement~\cite{Alva:2014gxa}.
Eq.~(\ref{eq:qaScattering}) is realization of the structure function approach to VBF~\cite{Han:1992hr}.
Formally, the $N\ell^\pm qq'$ channel can be recovered by performing the ACOT-like jet matching explicitly as in~\cite{Alva:2014gxa}
or evaluating the NLO in QED corrections to Eq.~(\ref{eq:qaScattering}).
For VBF, we impose the $\eta^{j},p_T^{j}$ cuts of~\cite{Alva:2014gxa}:
\begin{equation}
 \vert\eta^{j_\VBF}\vert < 4.5, \quad p_T^{j_\VBF} > 30\GeV.
\end{equation}
Collinear poles associated with $t$-channel $\ell$ exchange emerge in Eq.~(\ref{eq:qaScattering}) but are regulated by cuts in Eq.~(\ref{eq:LepCuts}).
The process at NLO in QCD is simulated by
\begin{verbatim}
> define q = u c d s b u~ c~ d~ s~ b~
> generate    q a > n2 mu q QED=3 QCD=0 [QCD]
> add process a q > n2 mu q QED=3 QCD=0 [QCD]
> output PP_VBF_NLO; launch;
\end{verbatim}

\section{Results}
\begin{table*}[!t]
\begin{tabular}{ c | c | c | c | c | c | c | c | c | c | c | c | c}
\hline \hline
$\sqrt{s}$ & \multicolumn{6}{c}{14 TeV} &\multicolumn{6}{c}{100 TeV} \tabularnewline\hline
\cline{2-9}
$m_N$ &\multicolumn{3}{c|}{500\GeV}&\multicolumn{3}{c|}{1 TeV}&\multicolumn{3}{c|}{500 GeV}&\multicolumn{3}{c}{1 TeV} \tabularnewline\hline
$\sigma ~/~ \vert V_{\ell N}\vert^2 ~$ [fb]  & ~LO~  & ~NLO~ & ~$K$~  & ~LO~  & ~NLO~ & ~$K$~   & ~LO~  & ~NLO~ & ~$K$~  & ~LO~  & ~NLO~ & ~$K$   
\tabularnewline\hline 
CC DY		& 52.8	& 61.1$^{+1.9\%}_{-1.6\%}$	& $1.16 $ & 2.96 & 3.40$^{+2.2\%}_{-2.4\%}$	& $1.15 $	
		& 674	& 804$^{+2.4\%}_{-3.4\%}$	& $1.19 $ & 80.8 & 93.5$^{+1.4\%}_{-1.6\%}$ 	& $1.16 $
\tabularnewline\hline
NC DY		& 30.4	& 35.2$^{+1.8\%}_{-1.5\%}$ 	& $1.16 $ & 1.56 & 1.81$^{+2.4\%}_{-2.5\%}$ 	& $1.16 $	
		& 537	& 638$^{+2.5\%}_{-3.6\%}$	& $1.19 $ & 55.9 & 64.4$^{+1.5\%}_{-1.7\%}$  	& $1.15 $     
\tabularnewline\hline
CC DY$+1j$	& 14.5	& 17.0$^{+3.2\%}_{-4.5\%}$ 	& $1.17 $ & 0.970 & 1.17$^{+4.0\%}_{-5.6\%}$ 	& $1.21 $	
		& 238	& 280$^{+2.1\%}_{-3.0\%}$ 	& $1.18 $ & 35.8 & 40.3$^{+2.0\%}_{-2.4\%}$	& $1.13 $
\tabularnewline\hline
GF$+0,1j$	& ~17.9~ & $\dots$ & $\dots$	& ~0.967~	& $\dots$ & $\dots$ & ~1,260~	& $\dots$ & $\dots$	& ~200~	&  $\dots$ & $\dots$     
\tabularnewline\hline
VBF		& 15.0	& 15.0$^{+7.8\%}_{-7.3\%}$ 	& $0.998 $ & 4.97 & 5.28$^{+6.3\%}_{-5.4\%}$ 	& $1.06 $	
		& 139	& 128$^{+12.3\%}_{-11.7\%}$ 	& $0.918 $ & 78.4 &  73.2$^{+10.0\%}_{-9.7\%}$	& $0.932 $
\tabularnewline\hline
\hline
\end{tabular}
\caption{LO and NLO heavy neutrino production rates, divided by active-heavy mixing $\vert V_{\ell N}\vert^2$, 
and scale dependence (\%) in $\sqrt{s} = 14$ and 100 TeV $pp$ collisions for representative heavy neutrino masses $m_N$.}
\label{tb:xSec}
\end{table*}

As a function of $m_N$, we present in Fig.~\ref{fig:xSec} the (a) 14 and (b) 100 TeV heavy $N$ production rates, divided by active-heavy mixing.
At NLO are the CC DY  (circle), NC DY (triangle), $N\ell^\pm+1j$ (diamond), and VBF (upside-down triangle) processes; at LO is GF (star).
The lower panel shows the NLO-to-LO ratio,
the so-called NLO $K$-factor:
\begin{equation}
 K^{NLO} \equiv \sigma^{\rm NLO} / \sigma^{\rm LO}.
\end{equation}
For select $m_N$, we summarize our results in Tb.~\ref{tb:xSec}.

For $m_N = 100-1000~(100-1500)\GeV$, NLO production rates for the DY channels at 14 (100) TeV span:
\begin{eqnarray}
 \DYCC	&:& {3.4\fb-16\pb	~\quad(25\fb-94\pb)},\\
 +1j	&:& {1.2\fb-2.1\pb	~\quad(12\fb-15\pb)},\\
 \DYNC	&:& {1.8\fb-23\pb	~\quad(16\fb-180\pb)},
\end{eqnarray}
with corresponding scale uncertainties:
\begin{eqnarray}
 \DYCC	&:& {\pm 1-5\%	~\quad(\pm1-11\%)},\\
 +1j	&:& {\pm 2-6\%	~\quad(\pm1-7\%)},\\
 \DYNC	&:& {\pm 1-5\%	~\quad(\pm1-13\%)},
\end{eqnarray}
and nearly identical $K$-factors:
\begin{eqnarray}
\DYCC, ~+1j, ~\text{NC} :\quad {1.15-1.25~(1.11-1.37)}.
\end{eqnarray}
The increase over LO rates is due to the opening of the $g\overset{(-)}{q}$  and $gg$ channels for the DY and $+1j$ processes, respectively.
Since the gluon PDF is largest at Bjorken-$x\sim m_N/\sqrt{s}\ll1$, the biggest change is at low $m_N$.
We find the that DY$+2j$ $K$-factors are consistent with high-mass SM DY in SHERPA~\cite{Gleisberg:2008ta}.
The modest size of these corrections validates our approach.

The VBF rate, uncertainty, and $K$-factor span
\begin{eqnarray}
 \sigma_{\VBF}			&:& {5.3-52\fb	~\quad(46-280\fb)},\\
 \delta\sigma_{\VBF}/\sigma	&:& {\pm5-11\%	~\quad(\pm9-14\%)},\\
 K_{\VBF} 			&:& {0.98-1.06	~\quad(0.90-0.96)}.
\end{eqnarray}
Due to collinear logarithmic enhancements, the VBF rate falls slower with $m_N$ than $s$-channel mechanisms.
At 14 (100) TeV, the VBF rate surpasses the CC DY rate at $m_N\approx {850~(1100)\GeV}.$
This somewhat differs from~\cite{Alva:2014gxa} and can be traced to the different $\gamma$-PDFs used:
at large (small) scales of $\tau = m_N^2/s$, the $q\gamma$ luminosity here is larger (smaller) than in~\cite{Alva:2014gxa},
leading to VBF overtaking the DY CC at smaller (larger) values of $m_N$.
However, present-day $\gamma$-PDF uncertainties are sizable~\cite{Ball:2013hta,Ababekri:2016kkj}. 

For all NLO processes, our scale dependence peaks at ${m_N = 100-200\GeV}$; 
it is attributed, in part, to the large gluon PDF uncertainty at small $x$.

For $m_N\geq 200\GeV$, the matched LO GF rate spans
\begin{eqnarray}
 \sigma_{\GF}	&:& {1.0\fb-0.1\pb	~\quad(55\fb-4.7\pb)}.
\end{eqnarray}
At 14 TeV, the rate is comparable to $N\ell+1j$ at NLO.
Though both obey $s$-channel scaling, the similarities are accidental and due to phase space cuts.
GF is roughly ${0.1-0.3\times}$ the CC DY rate.
At 100 TeV, the situation is qualitatively different:
Due to $gg$ luminosity increase at 100 TeV, which grows $\sim10\times$ more than the DY luminosity~\cite{Arkani-Hamed:2015vfh},
GF jumps to ${0.4-2\times}$ the CC DY rate, becoming the dominant  production mode for $m_N = {300-1500\GeV}$.
Beyond this $m_N$, VBF is largest. 
We observe that Higgs and $Z$ diagrams contribute about equally at large $m_N$.
Our matched results are consistent with the unmatched calculation of~\cite{Hessler:2014ssa}.

\subsection{NLO+PS Kinematics at 14 TeV}
We now consider the differential distribution for the processes in Fig.~\ref{fig:diagrams} but focus largely on the VBF channel.
The kinematics of heavy lepton production from DY currents at NLO and NLO+Leading Log(recoil) was studied in~\cite{Ruiz:2015zca}.
There, the differential NLO $K$-factors, defined as
\begin{equation}
 K_{\mathcal{O}}^{\rm NLO} \equiv \cfrac{d\sigma^{\rm NLO}/d\mathcal{O}}{d\sigma^{\rm LO}/d\mathcal{O}}
\end{equation}
for observable $\mathcal{O}$, were analytically shown to be flat in the leading regions of phase space.
In these regions, NLO contributions are dominated by soft initial-state radiation, which generically factorize for DY processes.
We confirm the flatness of $K_{\mathcal{O}}^{\rm NLOPS}$ for the DY channels, 
including for complex observables such as cluster mass in the $N\ell\rightarrow 3\ell \nu$ final state.

The phenomenology of the GF channel has not been previously studied.
It is beyond the scope of this investigation to do so here and will be presented elsewhere.

\begin{figure*}[!th]
  \subfigure[]{	\includegraphics[width=0.47\textwidth]{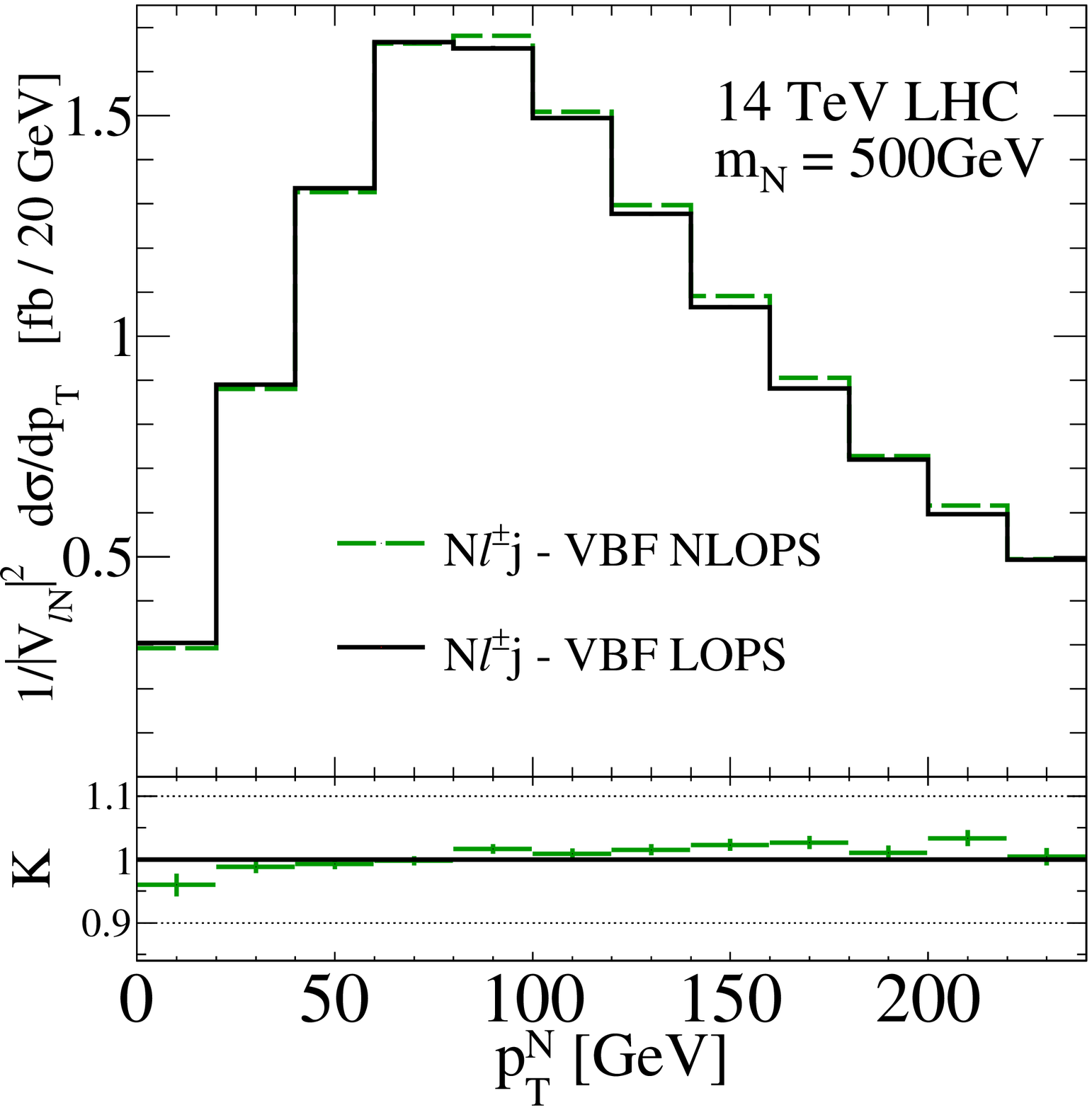}	\label{fig:vbf_pTN}	}
  \subfigure[]{	\includegraphics[width=0.47\textwidth]{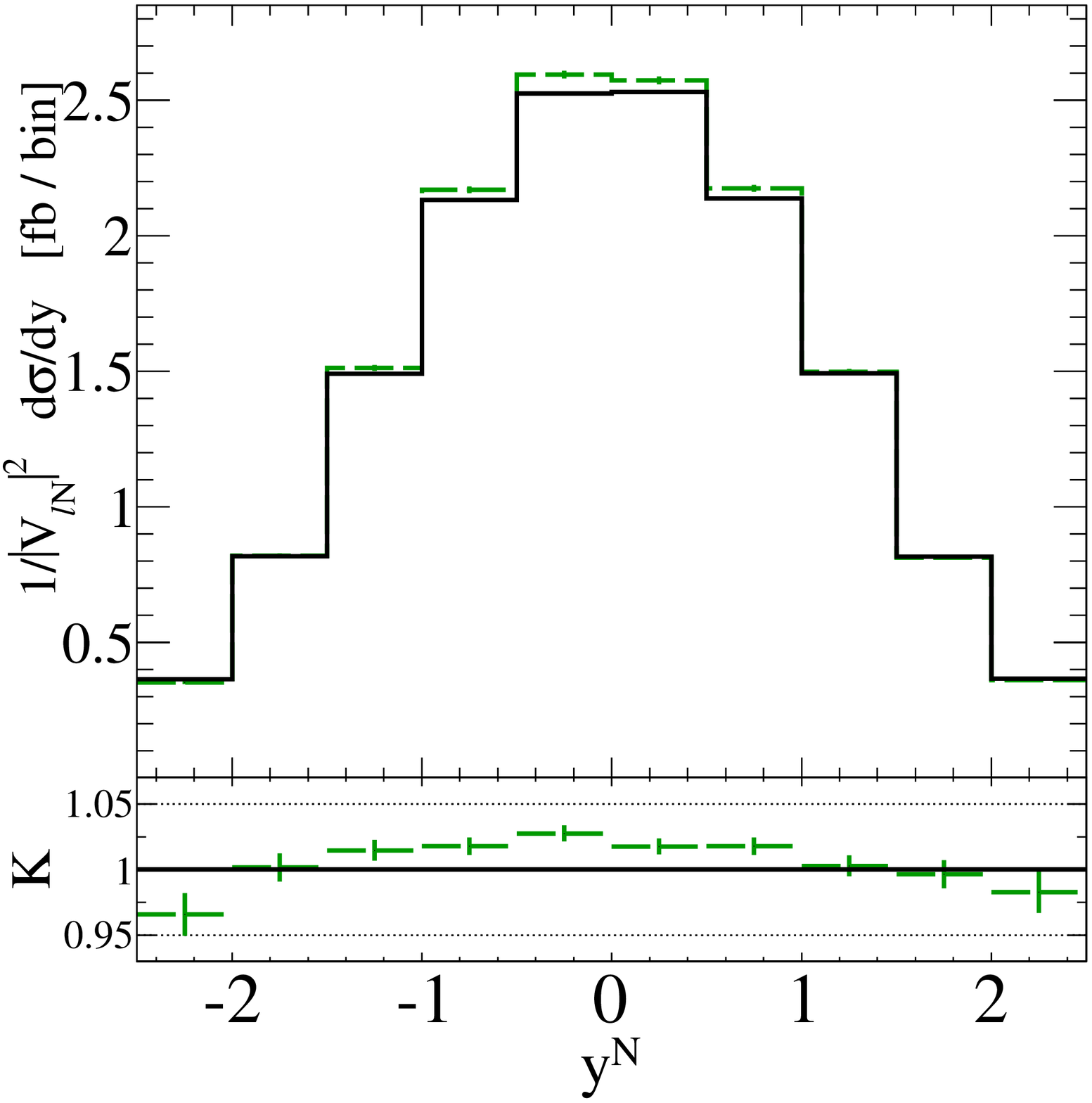}	\label{fig:vbf_yN}	}
  \\
  \subfigure[]{	\includegraphics[width=0.47\textwidth]{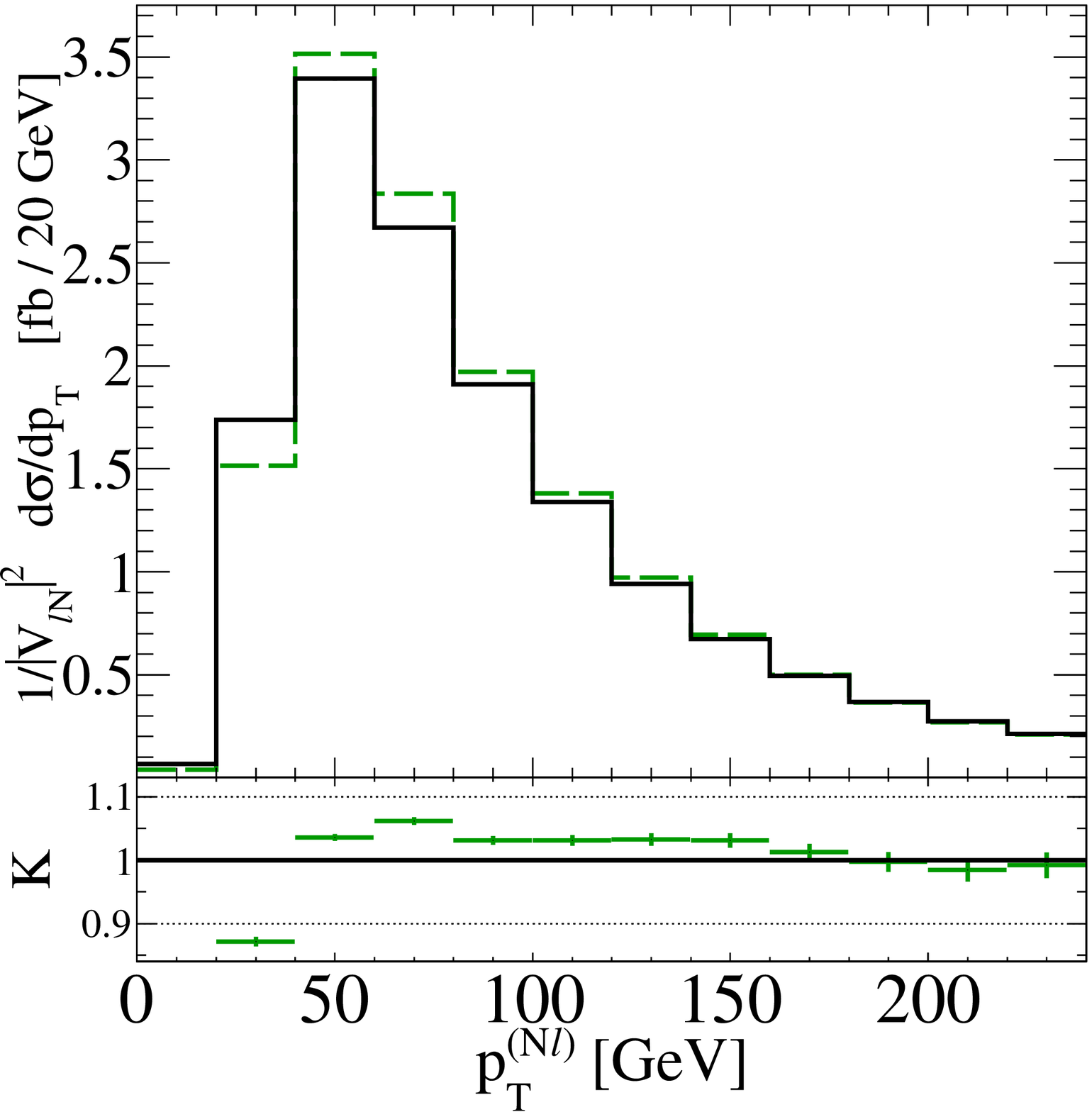}	\label{fig:vbf_pTNlsys}	}
  \subfigure[]{	\includegraphics[width=0.47\textwidth]{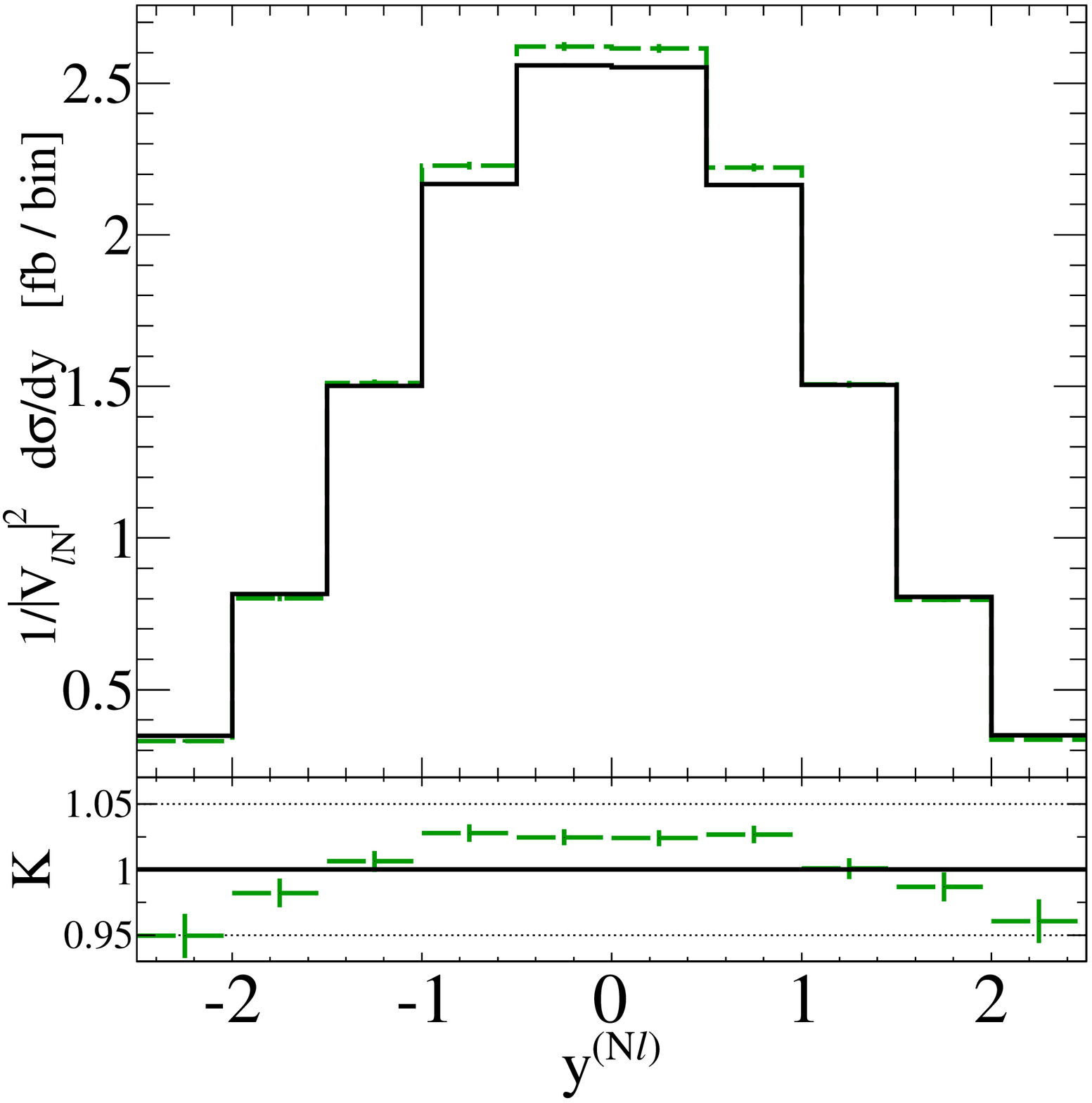}	\label{fig:vbf_yNlsys}	}
\caption{
Differential distributions with respect to (a,c) $p_T$ and (b,d) $y$ of (a,b) $N$ and (c,d) the $(N\ell)$ system at NLOPS (dash) and LOPS (solid) accuracy
in VBF at 14 TeV LHC for representative $m_N$ 500 GeV. Lower: Ratio of NLOPS and LOPS rates.
}
\label{fig:vbfDifferential}
\end{figure*}

At $\sqrt{s} = 14$ TeV and representative  neutrino mass $m_N = 500\GeV$,
the LO distributions for the $W\gamma$ fusion process was studied in Ref.~\cite{Alva:2014gxa}.
For the first time, we show in Fig.~\ref{fig:vbfDifferential} the NLO+PS (dash) and LO+PS (solid)-accurate distributions with respect to 
(a,c) $p_T$ and (b,d) rapidity $(y)$ of the (a,b) $N$- and (c,d) $(N\ell)$-systems. $K_{\mathcal{O}}^{\rm NLOPS}$ is shown in the lower panel.
For the two system, but particularly the $(N\ell)$-system, we observe a net migration at NLOPS of events from the lowest $p_T$ bins
resulting in $K^{\rm NLOPS}_{p_T}<1$ for these bins. At high $p_T$, $K^{\rm NLOPS}_{p_T}$ quickly converges to unity from above.
In the rapidity distributions, we observe a similar, but more pronounced, migration of events from large $y$ to smaller values,
consistent with shifts to larger $p_T$.
The charged lepton $p_T$ and $\eta$ distributions (not shown) demonstrate little sensitivity to $\mathcal{O}(\alpha_s)$ corrections.
However, as VBF is dominated by $\gamma\rightarrow \ell$ splittings~\cite{Alva:2014gxa}, 
one does not expect such sensitivity to QCD radiation until $\mathcal{O}(\alpha^2 \alpha_s)$.
Though numerically less significant, we find that the NLO corrections to the VBF distributions are qualitatively different than those of DY-like systems:
whereas differential $K$-factors for DY processes tend to remain flat and above unity,
QCD corrections for $W\gamma$ fusion tend to depopulate low-$p_T$/forward regions of phase space and populate high-$p_T$/central regions.
This results in $K$-factors both above and below unity.

\section{Summary and Conclusion}\label{sec:Conclusions}
The origin of light neutrino masses remains elusive.
Extended neutrino mass models predict the existence TeV-scale heavy neutrinos $N_i$ that may be discovered at current or future collider experiments.

We propose a systematic treatment of $N$ production mechanisms at hadron colliders, 
and provide instructions for building IRC-safe VBF and $N\ell^\pm+nj$ signal definitions.
The prescription remedies issues that have plagued past analyses,
and is applicable to a number of other SM and BSM processes.
We report modest NLO corrections, demonstrating the perturbative stability of our approach.
We present also the first NLOPS-accurate differential distributions for the $W\gamma$ VBF process.
We observe nontrivial differential $K$-factors below and above unity.

In a model-independent fashion,
we present for the first time a comparison of all leading single $N$ production modes at $\sqrt{s} = 14$ and $100$ TeV.
Fully differential results up to NLO in QCD accuracy are obtained through a MC tool chain linking FeynRules, \nloct, and \mgamc.
Associated model files are publicly available~\cite{nloFRModel}.
\\

\noindent 
\acknowledgements{
\textit{
Acknowledgements: D.~Alva, L.~Brenner, B.~Fuks, T.~Han, V.~Hirschi, J.~Rojo, S.~Pascoli, C.~Tamarit, C.~W\'eiland 
are thanked for discussions and readings of the manuscript.
S.~Kuttimalai and T.~Morgan are thanked for their numerical checks.
This work has been supported by Science and Technology Facilities Council (STFC),
the European Union’s Horizon 2020 research, and  innovation programme under the Marie Skłodowska-Curie grant agreements No 690575 and 674896.
OM and CD are Durham International Junior Research Fellow.
}}


\end{document}